\documentclass[11pt]{article}
\usepackage{epsf}
\begin{document}
% declarations for front matter
\begin{center}
\vspace*{20mm}
    {\huge Anisotropic Membranes\footnote{Research supported by the
       	Department of Energy U.S.A under Contract
       	No. DE-FG02-85ER40237, by Syracuse University, by Alexander
       	von Humboldt Stiftung, and by the Deutsche
       	Forschungsgemeinschaft.}}\\[2em] 
    {\large \em Mark Bowick\footnote{Talk given at Lattice 98, 
	Boulder Colorado, July 1998.}, Simon Catterall, Simeon Warner}\\ 
        {Physics Department, Syracuse University\\ 
	Syracuse, NY 13244--1130, U.S.A.}\\[1em] 
    {\large \em Gudmar Thorleifsson}\\
	{Fakult\"at f\"ur Physik, Universit\"at Bielefeld\\
	Bielefeld D--33615, Germany}\\[1em] 
    {\large \em Marco Falcioni}\\
        {Chemical Engineering Department, UCLA\\
	Los Angeles, CA 90095--1592 U.S.A.}\\[1em]
    {September 1998}\\[.5em] 
    {Syracuse University Preprint SU-4240-684}\\[1em]
\end{center}

\centerline{\bf Abstract}
\begin{quote}
We describe the statistical behavior of anisotropic crystalline membranes.
In particular we give the phase diagram and critical exponents 
for phantom membranes and discuss the generalization to self-avoiding membranes.
\end{quote}

\section{Introduction}
\label{intro}
\begin{figure}[htb]
\epsfxsize=4in
\centerline{\epsfbox{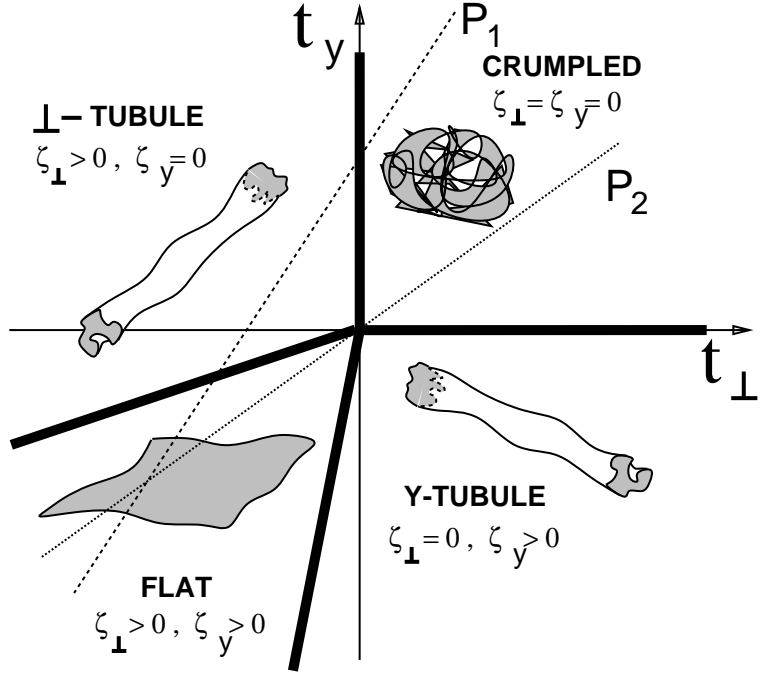}}
\caption{The mean field theory phase diagram for phantom
anisotropic membranes from \cite{RT95}.}
\label{phasedg}
\end{figure}

The statistical mechanics of phantom tethered membranes
with bending rigidity is quite well understood if the elastic and 
bending moduli are isotropic. There is a continuous {\em crumpling}
transition from the expected high-temperature {\em crumpled} phase
to a low-temperature {\em flat} phase that has no analogue in 
conventional spin systems \cite{revs,lh94}.
This flat phase is characterized by long-range orientational order 
of the membrane in the embedding space and large longitudinal (in-plane)
fluctuations. 
%The stability of the flat phase is due to the effective
%stiffening of the bending rigidity at long wavelength arising from the 
%nonlinear coupling to short wavelength thermal fluctuations.
Two important generalizations of this class of membranes arise by
incorporating {\em self-avoidance} and/or intrinsic {\em anisotropy}.
In \cite{RT95} Radzihovsky and Toner (RT) showed that intrinsic
anisotropy generally leads to a much richer phase diagram for {\em
phantom} (non-self-avoiding) tethered membranes.  Although anisotropy
is irrelevant in the flat and crumpled phases it leads to an entirely
new {\em tubular} phase separated by distinct continuous phase
transitions from both the conventional phases. This phase diagram is
shown in Fig.~\ref{phasedg}.

>From the applied point of view both isotropic and anisotropic
membranes should be important.  Isotropic membranes may be made by
suitable random polymerization of fluid vesicles.  On the other hand
polymerization in the presence of an applied electric field should in
principle yield an anisotropic membrane \cite{Ben97}.  Polymerized
membranes with in-plane tilt order have a natural
anisotropy.
The tubular phase is characterized by long-range orientational order
in one (extended) direction only {---} in the remaining transverse
direction the membrane is crumpled.  

\begin{figure}[htb]
\epsfxsize=4in
\centerline{\epsfbox{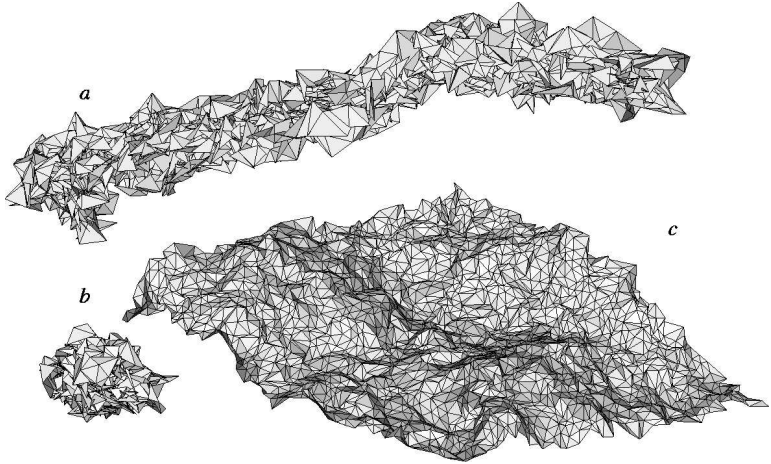}}
\caption{The three phases of anisotropic membranes:(a) tubular (b) crumpled 
and (c) flat.}
\label{tubules}
\end{figure}

The tubular phase thus resembles a rough sausage with a crumpled
cross-section.  This existence of this totally tubular phase was
confirmed by Monte Carlo simulations in \cite{BFT}. In this paper the
size (Flory) exponent $\nu_F$, describing the growth of the
tubule-diameter $R_G$ with internal system size (L), and the roughness
exponent $\zeta$, characterizing the height fluctuations along the
extended tubule axis, were also reported.  The results
$\nu_F=0.305(14)$ and $\zeta=0.895(60)$ were in rough qualitative
agreement with the theoretical predictions $\nu=\frac{1}{4}$ and
$\zeta=1$ of \cite{RT95}. More extensive subsequent simulations
\cite{BFT98} have improved these first results {--} the best current
results are $\nu_F=0.269(7)$ and $\zeta=0.859(40)$. Configurations
characteristic of the various phases are shown in Fig.~\ref{tubules}.

\section{Model}
\label{model}
A tethered membrane is described by a 2-dimensional regular
triangulated net, with the topology of a disk. Each node of the net
has six neighbors, except at the boundary. The Hamiltonian of the
system is
\begin{eqnarray}
 {\cal H}[{\rm \bf r}] & = & \sum_{\langle\sigma \sigma^\prime\rangle} \left|
 {\rm \bf r}_{\sigma} 
 - {\rm \bf r}_{\sigma^{\prime}} \right|^2  \nonumber\\
 & - & \kappa_1 {\sum_{\langle ab\rangle}}^{(x)} {\bf n}_a \cdot {\bf n}_b
 - \kappa_2 {\sum_{\langle ab \rangle}}^{(y)} {\bf n}_a \cdot {\bf n}_b \,.  
\label{eq:ham}
\end{eqnarray}
where ${\rm \bf r}_\sigma$ is the position in 3-dimensional space of
the node labelled $\sigma = (\sigma_x, \sigma_y)$.  The first sum runs
over all nearest-neighbor pairs (bonds) of the membrane, and is the
tethering potential. The second and third term are the bending
energies in the $x$ and $y$ intrinsic directions.  The bending energy
is a ferromagnetic interaction between the unit normals to the faces
of the membrane.  The strength of this interaction is anisotropic: if
two adjacent faces share a bond parallel to the $x$ direction the
coupling is $\kappa_1$; otherwise it is $\kappa_2$.

The canonical partition function for a membrane of fixed number of
nodes $N$ is
\begin{equation}
 Z \:=\: \int [\rm d{\bf r}] \: \delta ({\bf r}_{cm}) \:
 {\rm e}^{\textstyle - {\cal H}[\bf r] }.
\label{eq11}
\end{equation}

%\section{Numerical Methods}
%\label{methods}
The Hamiltonian of Eq.(\ref{eq:ham}) was simulated using Monte Carlo
methods for triangular lattices ranging from $25^2$ to $175^2$ nodes.
Further numerical details are given in \cite{BFT,BFT98}.
%\begin{figure}[htb]
%\vspace{9pt}
%\epsfxsize=2.8in 
%\epsfbox{fig2.eps}
%\caption{The phase diagram of an anisotropic tethered membrane.  The
% circles correspond to observed peaks in the specific heats $C_V^x$
% and $C_V^y$ (the filled ones are from larger lattices).  We performed
% simulations along the dotted lines with the cross indicating where we
% studied the tubular phase.}
%\label{mcpd}
%\end{figure}

\section{Global Phase Diagram}
\label{phases}
 
To determine the global phase diagram of anisotropic membranes we
explored the line $(\kappa_1,\kappa_2) = (3\kappa, \kappa)$.  We have
also simulated the lines $(\kappa,0)$ and $(2,\kappa)$ on smaller
lattices.  Distinct signatures of phase transitions were found in the
specific heats $C_V^i$ associated with the variance of the two bending
energy terms in the action $E_x$ and $E_y$;
\begin{equation}
C_V^i(\kappa) = \frac{\kappa^2}{L^2} \;
 \frac{\partial}{\partial \kappa} 
\langle E_i \rangle.
\end{equation}
\begin{figure}[t]
\epsfxsize=4in 
\centerline{\epsfbox{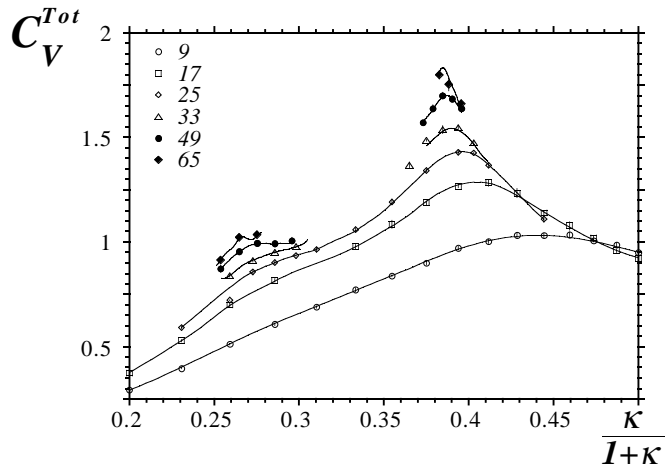}}
\caption{The value of the total specific heat
 $C_V$ for various lattice sizes. The interpolating lines are obtained using 
 multi-histogramming methods.}
\label{cvtot}
\end{figure}
In Fig.~\ref{cvtot} we show the total specific heat along the line 
$\kappa_1 = 3\kappa_2$ for lattice sizes up
to $175^2$.  There are two distinct peaks. We have confirmed that the
two peaks occur at {\it different} values of the bending rigidity,
signaling the existence of {\it two} distinct transitions.  
    
Performing a similar analysis along the other two lines in the
$(\kappa_1, \kappa_2)$ plane yields a phase diagram consistent with
that of Fig.~\ref{phasedg}.  This confirms the three phase scenario.
%This implies a three phase structure, the
%usual high-temperature crumpled and low temperature flat phases,
%together with the intermediate tubular phase predicted in \cite{RT95}.

\section{Tubular Exponents}
\label{tubulphase}

%To investigate the nature of this phase we have performed extensive
%simulations at the coupling $(\kappa_1=2,\kappa_2=0.4)$ on lattices
%up to $175^2$ in extent.  
A more detailed understanding of this tubular phase is obtained by
looking at the fluctuations of the zero-mode of the tubule height
$h_{rms}$, analogous to the height fluctuations of a flat membrane,
and the scaling of the width of the tubule $R^G_{\perp}$ \cite{BFT,BFT98}.
%\begin{figure}[htb]
%\vspace{9pt}
%\epsfxsize= 2.8in 
%\epsfbox{fig4.eps}
%\caption{The definition of the fluctuations of the zero-mode 
%$h_{rms}$ and the width of the tubule $R^G_{\perp}$.  
%The dotted line indicates the one-dimensional shape of the tubule,
%whereas the dashed one is the optimal straight line. } 
%\label{fig4}
%\end{figure}
These are expected to scale as $h_{rms} \sim L^{\zeta}$ and
${R^G_{\perp}} \sim L^{\nu_F}$.

We find $\zeta = 0.859(40)$ and $\nu_F = 0.269(7)$. The result for
$\nu_F$ is in excellent agreement with the analytic continuum
prediction of \cite{RT95} ($\nu_F=1/4$) but the roughness exponent is
well below the analytic prediction $\zeta=1$. Further studies are
under way \cite{BFT98}.
\begin{figure}[t]
\epsfxsize=4in 
\centerline{\epsfbox{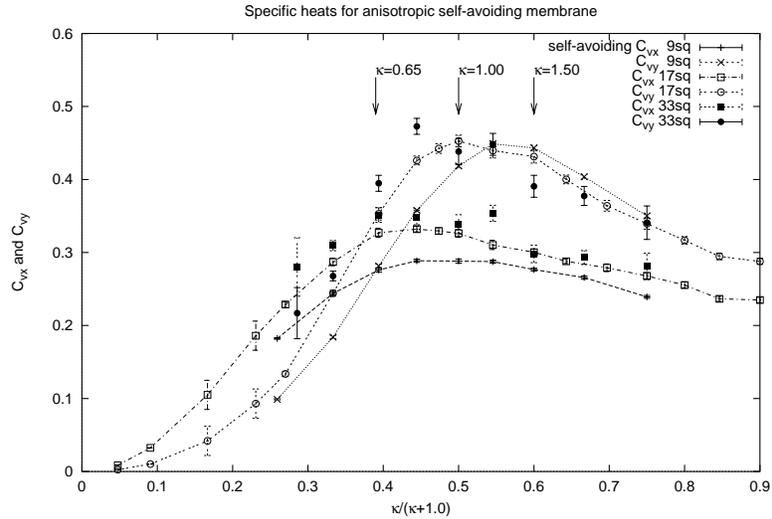}}
\caption{The specific heat $C_V$ for self-avoiding membranes.}
\label{sacv}
\end{figure}

The most challenging problem in the physics of tubules is the
incorporation of {\em self-avoidance}. Self-avoidance becomes relevant
below the upper-critical embedding dimension $d=11$, and here one can
develop a systematic $\epsilon$ expansion \cite{BG97}.  This
expansion can be improved and this is treated in the contribution to
this proceedings by Alex Travesset \cite{BT98,BTlat98}.  We are also
performing Monte Carlo simulations of this system.  Preliminary
results indicate a flat-to-tubular transition in the physical case of
three dimensions. The specific heat plot is shown in Fig.~\ref{sacv}.

\end{document}